\documentclass[10pt]{article}

\usepackage[utf8]{inputenc}
\usepackage{amsmath}
\usepackage{amsfonts}
\usepackage{amssymb}
\usepackage{authblk}
\usepackage{biblatex}
\usepackage[margin=1.375in]{geometry}
\usepackage{graphicx}
\usepackage{hyperref}
\usepackage{cleveref} 
\usepackage{xcolor}
\usepackage{xfrac}

\newcommand{\Metric}{\mathrm{Metric}}
\newcommand{\AltMetric}{\mathrm{Metric}'}
\newcommand{\optimum}{\mathrm{optimum}}
\DeclareMathOperator{\var}{var}
\DeclareMathOperator{\EE}{\mathbb{E}}
\DeclareMathOperator{\PP}{\mathbb{P}}

\Crefname{equation}{Eq.}{Eqs.}
\creflabelformat{equation}{#2\textup{#1}#3}

\title{The Bias-Variance Tradeoff in Long-Term Experimentation}
\author[1]{Daniel Ting}
\author[2]{Kenneth Hung}
\affil[1]{Meta \\ {\tt dting@meta.com}}
\affil[2]{Central Applied Science, Meta \\ {\tt kenhung@meta.com}}

\date{November 4, 2025}

\begin{document}

\maketitle

\section{Introduction}

As we exhaust methods that reduces variance without introducing bias \cite{ting2023limits}, reducing variance in experiments often requires accepting some bias, using methods like winsorization \cite{wang2019measuring} or surrogate metrics \cite{duan2021online}. While this bias-variance tradeoff can be optimized for individual experiments, bias may accumulate over time, raising concerns for long-term optimization. We analyze whether bias is ever acceptable when it can accumulate, and show that a bias-variance tradeoff persists in long-term settings. Improving signal-to-noise remains beneficial, even if it introduces bias. This implies we should shift from thinking there is a single ``correct'', unbiased metric to thinking about how to make the best estimates and decisions when better precision can be achieved at the expense of bias.

Furthermore, our model adds nuance to previous findings  \cite{azevedo2020abtesting,sudijono2024optimizing} that suggest less stringent launch criterion leads to improved gains. We show while this is beneficial when the system is far from the optimum, more stringent launch criterion is preferable as the system matures.

\section{Experiment-based optimization as stochastic gradient descent}

We study this problem by treating our long term optimization problem like a stochastic gradient descent (SGD) procedure \cite{welling2011bayesian}. Each experiment can be treated as a noisy measurement of the lift, so each launch is a noisy step towards or away from the global optimum. This sets up a stochastic differential equation (SDE) that we can solve. By injecting bias and variance into the procedure, we can study whether the optimality gap increases or decreases.

For simplicity, we consider a problem where we optimize a metric by changing a parameter. When the parameter is $x_t$ at time $t$, the observed metric is $\Metric(x_t)$. This contains some sampling noise and has expectation $\EE \Metric(x_t)$, an objective that we wish to optimize. Without loss of generality, we can assume $\EE \Metric(x)$ attains the optimum at $x = 0$. Since at the optimum, the true objective has zero gradient, we can express the objective as
\[
    \EE \Metric(x_t) = \optimum - \frac{1}{2}r x_t^2,
\]
where $r$ captures the scaling of the effect of the parameter $x$ on the outcome.

At time $t$, with probability $\sfrac{1}{2}$ we propose to change the parameter $x_t$ to $x_t - 1$ and with probability $\sfrac{1}{2}$ we propose changing it to $x_t + 1$. A change from $x_t$ to $x_t - 1$ will change the objective by
\begin{equation}
\label{eq:true-effect}
    \EE \Metric(x_t - 1) - \EE \Metric(x_t) \approx - \nabla \EE \Metric(x_t) = rx_t,
\end{equation}
which we can measure via an experiment. For simplicity, we assume all experiments to have standard error $\sigma$. The experiment is then launched if the result is statistically positive, so the probability of changing $x_t$ to $x_t - 1$ is
\[
    \PP(Z + rx_t / \sigma > c) = \bar\Phi(c - rx_t / \sigma),
\]
where $Z$ is a standard normal variable, $\bar\Phi$ is its survival function (or one minus its cumulative distribution function), and $c$ is the hypothesis testing threshold for a $z$-score at level $\alpha$, e.g.\ $c = 1.64$ for $90\%$ confidence. Similar, a change from $x_t$ to $x_t + 1$ is launched with probability $\bar\Phi(c + rx_t / \sigma)$. In other words, given the parameter is $x_t$ at time $t$, the parameter at time $t+1$ will be given by
\[
    X_{t+1} = \begin{cases}
        x_t - 1 & \text{with probability } \frac{1}{2}\bar\Phi(c - rx_t / \sigma), \\
        x_t + 1 & \text{with probability } \frac{1}{2}\bar\Phi(c + rx_t / \sigma), \\
        x_t & \text{otherwise.}
    \end{cases}
\]

Consider the case where it is difficult to detect a treatment effect, and the scale of noise is bigger than the effect size, i.e.\ $\sigma \gg |rx_t|$. The parameter $X_{t+1}$ will satisfy
\begin{align}
    \EE [X_{t+1} - x_t] &= -\frac{1}{2}\bar\Phi(c + rx_t / \sigma) + \frac{1}{2}\bar\Phi(c - rx_t / \sigma) 
    \approx rx_t \phi(c) / \sigma, \label{eq:drift} \\
    \var(X_{t+1} - x_t) &= \EE\left[(X_{t+1} - x_t)^2\right] - \EE [X_{t+1} - x_t]^2 \nonumber \\
    &\approx \frac{1}{2}\bar\Phi(c - rx_t / \sigma) + \frac{1}{2}\bar\Phi(c + rx_t / \sigma) - o(r x_t / \sigma) \nonumber \\
    &\approx \alpha. \label{eq:diff}
\end{align}

After rescaling time to obtain a continuous time process, these define the drift and diffusion terms of an SDE:
\begin{equation}
\label{eq:sde}
    dX_t = -\frac{r\phi(c)}{\sigma} x_t \,dt + \sqrt{\alpha} \,dW_t = -\theta x_t \,dt + \sqrt{\alpha} \,dW_t,
\end{equation}
where $\theta = r\phi(c) / \sigma$.

This is a well-known SDE for the Ornstein--Uhlenbeck process \cite{oksendal2003stochastic}, i.e.\ a mean reverting walk. This is a particularly nice SDE cause it has a closed form solution for its distribution at every time point. The solution and limit distribution are  given by
\begin{equation}
\label{eq:sde-solution}
    X_t \sim N\left(x_0 e^{-\theta t}, \frac{\alpha}{2\theta} \left(1 - e^{-2\theta t}\right)\right),
    \qquad
    X_\infty \sim N(0, \alpha / 2\theta)
\end{equation}
Correspondingly, the objectives at time $t$ and as $t \to \infty$ are
\begin{align}
    \EE \Metric(X_t) &= \optimum - \frac{1}{2} r \left(x_0 e^{-2\theta t} + \frac{\alpha}{2\theta} \left(1 - e^{-2\theta t}\right)\right) \label{eq:metric} \\
    \EE \Metric(X_\infty) &= \optimum - \frac{1}{4} \frac{r \alpha}{\theta} \label{eq:metric-inf}
\end{align}

\subsection{Interpreting the SDE solution}

\begin{figure}
    \centering
    \includegraphics[height=3in]{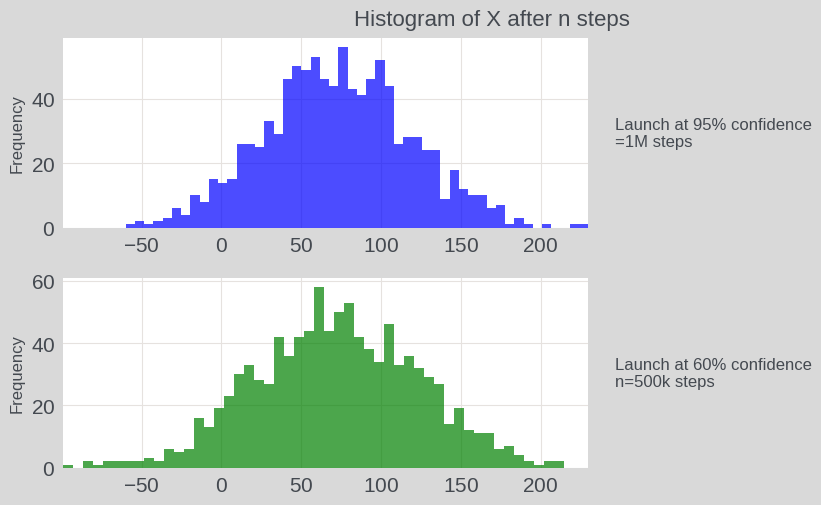}
    \caption{Distribution of the underlying state $X_n$ starting at $X_0=100$. Launches using statistical significance for a one-sided test at the $5\%$ level takes twice as long to converge as a one-sided test at the $40\%$ level.}
    \label{fig:convergence-speed}
\end{figure}

This equation already describes interesting properties of the stochastic optimization of systems through experimentation and highlight the importance of reducing the variance of lift estimates.
\begin{itemize}
    \item {\bf Longer convergence.} Increasing $\sigma$ to $\sigma'$ means it takes $(\sigma'/\sigma)$ times longer to reach the stationary distribution. In other words, you need to run $(\sigma'/\sigma)$ times more experiments to reach the same gains.
    \item {\bf Suboptimal final state.} A large standard deviation $\sigma$ in the lift measurement directly reduces the expected value of the target metric.
    \item {\bf Less stringent launch criterion yields faster convergence.} While larger $\sigma$ slows down convergence, smaller decision thresholds $c > 0$ increase $\phi(c)$ and $\theta$, which speeds up convergence as shown in \Cref{fig:convergence-speed}.
    \item {\bf More stringent launch criterion yields better final state.} For the optimality gap in \Cref{eq:metric-inf}, we see that
    \[
        \frac{1}{4} \frac{r \alpha}{\theta} = \frac{\sigma}{4} \frac{\bar\Phi(c)}{\phi(c)}.
    \]
    The ratio $\bar\Phi(c) / \phi(c)$ is the Mills's ratio for a Gaussian distribution and is $(1 + o(c)) (1/c)$ as $c \to \infty$.
\end{itemize}

While the first point is completely unsurprising, the second and fourth suggest that as the system becomes more and more optimized, experiments need to be run with higher power and more stringent launch criterion to reach a better final state!

Perhaps counterintuitively, the third and fourth points fall in line with previous suggestions to relax the $p$-value threshold in \cite{azevedo2020abtesting,sudijono2024optimizing} --- in a less mature system farther from the stationary state, effect sizes are bigger and a relaxed $p$-value threshold helps to achieve faster improvements; while in a mature system closer to the stationary state, effect sizes are smaller and a more stringent $p$-value threshold is preferred.

\section{Impact of bias}

But what happens when you induce bias? For simplicity, we consider two scenarios:
\begin{itemize}
    \item {\bf Bias in the updates.} The bias shrinks the experiment estimand toward zero by a factor of $\gamma_r$, in exchange for shrinking the standard error by a factor of $\gamma_\sigma$.
    \item {\bf Bias in the objective.} The metric is a biased measurement on the true estimand.
\end{itemize}
In the first scenario, we should still converge to a limiting distribution with the right mean, while in the second scenario we assume the bias actually changes the mean of the limiting distribution.

\subsection{Bias in the updates}
\label{sec:biased-updates}

In the first scenario, the true effect in \Cref{eq:true-effect} becomes $rx_t / \gamma_r$ and the standard error becomes $\sigma / \gamma_\sigma$. The SDE (\Cref{eq:sde}) can be updated by replacing $r$ with $r / \gamma_r$ and $\theta$ with $\frac{\gamma_\sigma}{\gamma_r} \theta$. By plugging this into the solution (\Cref{eq:sde-solution}), we can see that this scenario affects the solutions in how fast it converges, faster convergence is achieved when $\gamma_r < \gamma_\sigma$. In other words, it improves all aspects of the optimization if the signal-to-noise ratio improves so that $\frac{r / \gamma_r}{\sigma / \gamma_\sigma} > \frac{r}{\sigma}$. We will denote the improvement in signal-to-noise ratio as $\gamma$, i.e.\ $\gamma = \gamma_\sigma / \gamma_r$.

Furthermore, plugging this in \Cref{eq:metric,eq:metric-inf}, the new expected target metric is
\[
    \EE \Metric(X_\infty) = \optimum - \frac{1}{4} \frac{r \alpha}{\gamma \theta}.
\]

The key takeaway is that \emph{any} improvement in the signal-to-noise ratio is beneficial, even at the expense of biased experiment results when the bias just shrinks the effect size.

\subsection{Bias in the objective}

Now consider a metric that is a biased measurement of the true estimand, e.g.\ when the estimand cannot be directly measured or if one uses a surrogate metric. This alternative metric induces another objective with the corresponding Ornstein--Uhlenbeck dynamics
\[
    \EE \AltMetric(x_t) = \optimum - \frac{1}{2} r'(x - \mu)^2,
\]
where the optimizer of the objective is not zero but some value $\mu$. Suppose we can measure the changes in this alternative metric with standard error $\sigma'$. Using the same notation in \Cref{sec:biased-updates} where $\gamma$ stands for the improvement in the signal-to-noise ratio $\frac{r'}{\sigma'} / \frac{r}{\sigma}$, the solution becomes
\[
    X_t \sim N\left(x_0 e^{-\gamma \theta t} + \mu(1 - e^{-\gamma\theta t}), \frac{\alpha}{2\gamma\theta} \left(1 - e^{-2\gamma\theta t}\right)\right)
\]
and the true objective is
\[
    \EE \Metric(X_\infty) = \optimum - \frac{1}{2} r \left(\mu^2 + \frac{\alpha}{2 \gamma\theta}\right).
\]

Here, the final state improves if $\mu^2 < \frac{\alpha}{2\theta} (1 - \gamma^{-1})$. This shows there is a bias-variance tradeoff in this scenario as well and that benefits rely only on improvements to the signal-to-noise ratio. 

Furthermore, we note the relative importance of bias. While increasing bias has a quadratic impact on the final objective, increase the noise-to-signal ratio only has a linear effect. This suggests as systems become increasingly optimized, that surrogate metrics may need to be designed more carefully and favor reducing bias over reducing variance. How to empirically choose this tradeoff is still an open question.

\section{Discussion}

Methods such as winsorization introduce bias for the sample mean, akin to the attenuation bias in the scenario in \Cref{sec:biased-updates} as the same bias is introduced to both the test and control groups. Our demonstration provides some justification for these long-term bias-variance tradeoffs.

We demonstrate there are bias-variance tradeoffs in long-term experimentation. While bias in the objective can be considered relatively more important than variance since the expected value of the objective at $X_\infty$ is quadratic in the bias $\mu$ and linear in the standard error $\sigma$,
the notion that ``unbiasedness means correctness'' is incomplete. 

\printbibliography

\end{document}